\documentclass[prd, aps, superscriptaddress, preprintnumbers, twocolumn, floatfix, nofootinbib]{revtex4}

\usepackage{amsfonts}
\usepackage{amsmath}
\usepackage{amssymb}
\usepackage{bm}
\usepackage{dcolumn}
\usepackage{siunitx}
\usepackage{epsfig}
\usepackage[latin1]{inputenc}
\usepackage{latexsym}
\usepackage{rotating}
\usepackage{hyperref}

\newcommand\be{\begin{equation}}
\newcommand\ba{\begin{eqnarray}}
\newcommand\ee{\end{equation}}
\newcommand\ea{\end{eqnarray}}

\begin{document}

\title {Searching for Cosmic Strings in CMB Anisotropy Maps using Wavelets and Curvelets}

\author{Lukas Hergt}
\email{hergtl@phys.ethz.ch}
\affiliation{ETH Zurich, Department of Physics, Wolfgang-Pauli-Strasse 27, 8093 Zurich, Switzerland} 

\author{Adam Amara}
\email{adam.amara@phys.ethz.ch}
\affiliation{ETH Zurich, Department of Physics, Wolfgang-Pauli-Strasse 27, 8093 Zurich, Switzerland} 

\author{Robert Brandenberger}
\email{rhb@physics.mcgill.ca}
\affiliation{Physics Department, McGill University, Montreal, QC, H3A 2T8, Canada, and \\Institute for Theoretical Studies,
ETH Z\"urich, CH-8092 Z\"urich, Switzerland}

\author{Tomasz Kacprzak}
\email{tomasz.kacprzak@phys.ethz.ch}
\affiliation{ETH Zurich, Department of Physics, Wolfgang-Pauli-Strasse 27, 8093 Zurich, Switzerland} 

\author{Alexandre R\'efr\'egier}
\email{alexandre.refregier@phys.ethz.ch}
\affiliation{ETH Zurich, Department of Physics, Wolfgang-Pauli-Strasse 27, 8093 Zurich, Switzerland} 

\date{\today}

%%%%%%%%%%%%%%%%%%%%%%%%%%%%%%%%%%%%%%%%%%%%%%%%%%%%%%%%%%%%%%%%%%%%%%%%%%%%%%%%%%%%%%%%%%%%%%
\begin{abstract}

We use wavelet and curvelet transforms to extract signals of
cosmic strings from simulated cosmic microwave background (CMB) 
temperature anisotropy maps, and to study the limits on the
cosmic string tension which various ongoing CMB
temperature anisotropy experiments will be able to achieve.
We construct sky maps with size and angular resolution
corresponding to various experiments. These maps contain the
signals of a scaling solution of long string segments with
a given string tension $G \mu$, the contribution of the
dominant Gaussian primordial cosmological fluctuations,
and pixel by pixel white noise with an amplitude corresponding
to the instrumental noise of the various experiments. In the case that we include white noise, we find
that using curvelets we obtain lower bounds on the string tension than with wavelets.
For maps with Planck specification, we obtain bounds comparable to what was obtained by
the Planck collaboration \cite{PlanckCS}. Experiments
with better angular resolution such as the South Pole
Telescope third generation (SPT-3G) survey
will be able to yield stronger limits. For maps with a 
specification of SPT-3G  we find that string signals will be
visible down to a string tension of $G \mu = 1.4 \times 10^{-7}$.

\end{abstract}
%%%%%%%%%%%%%%%%%%%%%%%%%%%%%%%%%%%%%%%%%%%%%%%%%%%%%%%%%%%%%%%%%%%%%%%%%%%%%%%%%%%%%%%%%%%%%%

\maketitle

%%%%%%%%%%%%%%%%%%%%%%%%%%%%%%%%%%%%%%%%%%%%%%%%%%%%%%%%%%%%%%%%%%%%%%%%%%%%%%%%%%%%%%%%%%%%%%
\section{Introduction} 

Cosmic strings are linear topological defects which exist as solutions in a large
class of particle physics models beyond the Standard Model. Cosmic strings
have common features with line defects in crystals and with vortex lines in
superconductors and superfluids. Cosmic strings are lines of trapped energy
density, and as such they have gravitational effects on space-time which lead
to cosmological signals (for reviews on cosmic strings see e.g. \cite{VS, HK, RHBCSrev}).

If nature is described by a particle physics model which has cosmic string solutions,
then a network of strings inevitably forms during a phase transition 
in the early universe and persists to
the present time. This is a causality argument which is originally due to Kibble
\cite{Kibble}. Hence, the gravitational effects of the strings are unavoidable.
The gravitational effects of strings are characterized by a single number, namely
the tension $\mu$ (or $G \mu$ in terms of a dimensionless variable, where $G$
is Newton's gravitational constant). Since cosmic strings are relativistic, the
tension equals the energy per unit length. The tension is given by the energy
scale $\eta$ of the particle physics model
\be
\mu \, \simeq \, \eta^2 \, .
\ee
Hence, the gravitational effects of cosmic strings increase as the energy
scale of the particle physics model increases. Searching for cosmological
signatures of strings is hence a way to test particle physics models 
``from top down" (the effects are larger the higher the value of $\eta$ is)
and is complementary to accelerator investigations which probe particle
physics models ``from bottom up" (the effects are larger the lower $\eta$ is -
see e.g. \cite{RHBrecentRev} for an elaboration on this point).

The network of cosmic strings consists of infinite strings with a
mean curvature radius and separation of $\xi$, and a distribution of
string loops with radii $R < t$, where $t$ is the time. Causality tells
us that
\be
\xi(t) \, \leq \, t \, 
\ee
at all times after the phase transition during which the string network
forms \cite{Kibble}, including the present time. If the strings form
during a typical symmetry breaking phase transition, then, 
at the time of formation, the network of cosmic strings is very dense,
i.e. $\xi(t) \ll t$. The late time behaviour of $\xi(t)$ can be obtained
from a Blotzmann equation which describes how the infinite string
network loses energy by splitting off string loops. The result of
solving this Boltzmann equation is \cite{VS, HK, RHBCSrev}
\be
\xi(t) \, \sim \, t 
\ee
at all late times. Thus, the network of strings becomes {\it scale-invariant}
in the sense that the statistical properties do not depend on time if all
lengths are scaled to the horizon $t$. The network thus consists of a few
infinite strings crossing each horizon volume, the strings having typical
curvature radius proportional to $t$, and a distribution of loops. In
this article we will focus on the signals of the infinite strings since they
lead to signatures with particular geometrical patterns in position space maps.

The gravitational signatures of strings are due to the fact that space
perpendicular to a string is conical with deficit angle \cite{deficit}
\be
\alpha \, = \, 8 \pi G \mu \, .
\ee
This deficit angle is visible up to a horizon distance in perpendicular
direction from the string, where it rapidly decreases to zero \cite{Joao}.

The conical structure of space perpendicular to a long string leads to
the {\it Kaiser-Stebbins} effect \cite{KS}: radiation passing on the two
sides of a moving string experiences a lensing Doppler shift of magnitude
\be
\frac{\Delta T}{T} \, = \, 8 \pi G \mu v_s \gamma_s \, ,
\ee
where $v_s$ is the string velocity and $\gamma_s$ is the
associated relativistic $\gamma$ factor. This effect leads to
an interesting signal of strings in cosmic microwave background (CMB)
temperature anisotropy maps:  strings will lead to a network of
line discontinuities which are caused by the lensing of the CMB
photons by strings which the photons pass while travelling from the
time $t_{rec}$ of recombination until today $t = t_0$. The characteristic
length scale of these lines is given by the comoving horizon at
the time $t$ when the photons pass by the string (e.g. about one degree
in the sky for strings close to $t_{rec}$), and the depth of the
signal is also comparable to this scale.

The density perturbations due to strings cannot be the dominant source
of fluctuations in the Universe (as was initially hoped in the 1980s - see e.g.
\cite{Zel, Vil, TB}). The reason is that fluctuations from strings are incoherent
and active and do not lead to acoustic oscillations in the CMB angular
power spectrum \cite{noacoustic}. In fact, the observation of these
oscillations provides a tight upper bound 
on the cosmic string tension \cite{Dvorkin, PlanckCS, CMBlimits}
\be
 G \mu \, \lesssim \, 10^{-7} \, 
\ee
 (the precise coefficient depends on the specific cosmic string
evolution simulation used).
Cosmic strings are hence only a supplementary source of fluctuations.
However, they lead to characteristic patterns in a number of observational
windows (see e.g. \cite{RHBCSrevNew}).

Most of the signatures of long cosmic strings are due to the fact that
moving strings present at time $t$ generate regions of twice the background density in
their wake \cite{wake}. These {\it string wakes} are nonlinear fluctuations of comoving
dimension
\be
z(t) \bigl[ t, v_s \gamma_s t, 4 \pi G \mu v_s \gamma_s t \bigr] \, ,
\ee
where $z(t)$ is the redshift at time $t$.
Wakes lead to planar overdensities of galaxies and hence to direct B-mode CMB polarization
\cite{Holder1} - rectangles in the sky with a uniform polarization direction and a
linearly growing amplitude. Wakes at high redshift also lead to thin slices in 21cm redshift
surveys where there is extra absorption \cite{Holder2}. The signatures of wakes
are more prominent at high redshifts than at low redshifts since the wakes get
disrupted by the dominant Gaussian perturbations at late times (see \cite{Disrael}
for a study of wake disruption).  

The overall angular power spectra or Fourier spectra of cosmic string signals is
scale-invariant and hence hard to distinguish from the effects of the dominant
Gaussian fluctuations (the fluctuations produced by the dominant source, e.g.
the $\Lambda$CDM fluctuations). This is shown e.g. in the computation of the angular
power spectrum of B-mode CMB polarization \cite{Salton} which yields a result
which is degenerate in shape with the B-mode polarization induced by lensing of
the E-mode polarization. Hence, to be able to efficiently search for strings it
is more promising to analyze position space maps.

We will here explore the power of wavelets and curvelets (see e.g \cite{wavelet} for
an introduction to wavelet analysis, and \cite{curvelet2000} for an introduction to curvelets)
to find signatures of cosmic strings in CMB temperature maps. There have been
some initial efforts at searching for strings in position space CMB temperature maps
in \cite{Smoot} and \cite{Wright}, the latter using a matched filtering analysis. Later on,
the power of the Canny Edge Detection algorithm \cite{Canny} to detect strings was
explored in \cite{Danos}. The analysis of \cite{Danos} showed that the Canny
algorithm might allow the bounds on the cosmic string tension to be reduced by close
to a factor of $10$ compared to what can be obtained in Fourier space when considering
$10^\circ \times 10^\circ$ degree maps of the sky at $^\prime$ resolution, specifications
chosen to correspond to the South Pole Telescope \cite{SPT} and Atacama Cosmology
Telescope \cite{ACT} projects \footnote{Note, however, that instrumental noise
effects were not considered in \cite{Danos}.}.

The Canny algorithm looks for edges in temperature maps where the gradient of the  
temperature is large and where the direction of the gradient does not change much.
Such an algorithm is suitable to find the Kaiser-Stebbins edges in CMB maps produced
by strings, but it turns out that the long edges get disrupted by the Gaussain noise,
and that the algorithm is hence not optimal. We are here searching for an algorithm
which allows the effects of the long strings to stand out even in the presence of
Gaussian noise, and we here show that wavelet and curvelet analysis can provide
a method.

In the following, we first describe the CMB temperature maps which we have
constructed. They contain contributions of a scaling network of
long cosmic string segments, {\it Gaussian noise},
by which we mean the Gaussian fluctuations which the standard $\Lambda$CDM model
would provide, and instrumental noise.  Our basic maps cover $12.8^\circ \times 12.8^\circ$ 
sections of the sky at $1.5^\prime$ resolution. By combining such maps and by
smoothing out the resulting maps we can construct maps relevant to the
characteristics of various ongoing experiments such as Planck \cite{Planck}
or the South Pole Telescope (SPT) project. 
We then introduce the wavelet and curvelet transforms which we use,
and describe their application to the temperature maps. We compare the
potential of various experiments to detect cosmic string signals, and
we find that the SPT-3G experiment \cite{SPT3G} will be able to identify string 
signals down to a string tension of $G \mu = 1.4 \times 10^{-7}$, a value comparable 
to the current best limits \cite{Dvorkin, PlanckCS}
which are derived from fitting the angular power spectrum of the CMB. Note that
in the absence of instrumental noise, signals are visible by eye down to a tension of $G \mu = 5\times 10^{-8}$, and that the
signals can be picked out by a simple histogram statistic down to a value of
$G \mu = 3 \times 10^{-8}$, a significantly better performance than what can be
achieved using the Canny algorithm. With instrumental noise effects included
our limit from maps of SPT-3G specification is $G \mu < 1.4 \times 10^{-7}$. 

Searches for the non-Gaussian signals of cosmic strings were also performed by
the Planck collaboration \cite{PlanckCS} using Minkowski functionals and steerable
wavelets (see also \cite{Starck} for a discussion of the method and \cite{Application} for initial applications to the search for cosmic strings in CMB maps. See also \cite{othermethods} for other methods to search for non-Gaussian signals of strings). 
The constraints obtained from these analyses are comparable to the ones
we obtain for simulations with the size, angular resolution and noise level of the
Planck maps. Our work is complementary to the analysis by the Planck collaboration.
The analysis of \cite{PlanckCS} is based on cosmic string temperature maps
generated from the numerical string network simulations of \cite{Ringeval} \footnote{See
also \cite{CSsimuls} for other cosmic string evolution simulations.}. Hence, these
maps contain both the signals from long strings and from loops, and they take into
account the fact that the long string segments can have small-scale structure. 
On the other hand, the analysis is subject to numerical uncertainties in the
simulations. Our analysis is based on toy model string maps produced by
long string segments without small-scale structure. The signals we find are
hence due to the purely stringy nature of the maps. Our approach also allows
us to vary the parameters which describe the cosmic string network. It is
satisfying that our results for maps with Planck specification match with those
obtained in \cite{PlanckCS}.

Note that, following the usual conventions in the cosmic string literature, we are using
natural units in which the speed of light $c$, Planck's constant and the Boltzmann
constant are set to $1$. 

\section{Simulations}

For this work we have produced new CMB temperature maps which
are the superposition of three maps, one due to Gaussian noise, 
the second due to cosmic strings, and the third due to instrumental noise. Our basic 
maps have an angular scale of $12.8^\circ \times 12.8^\circ$ and an angular resolution
of $1.5^\prime$, yielding maps with $512 \times 512$ pixels. Further down, these properties and the pixel by pixel white noise will be matched to various ongoing experiments.

\begin{figure*}[p]
\begin{center}
\includegraphics[height=0.87\textheight]{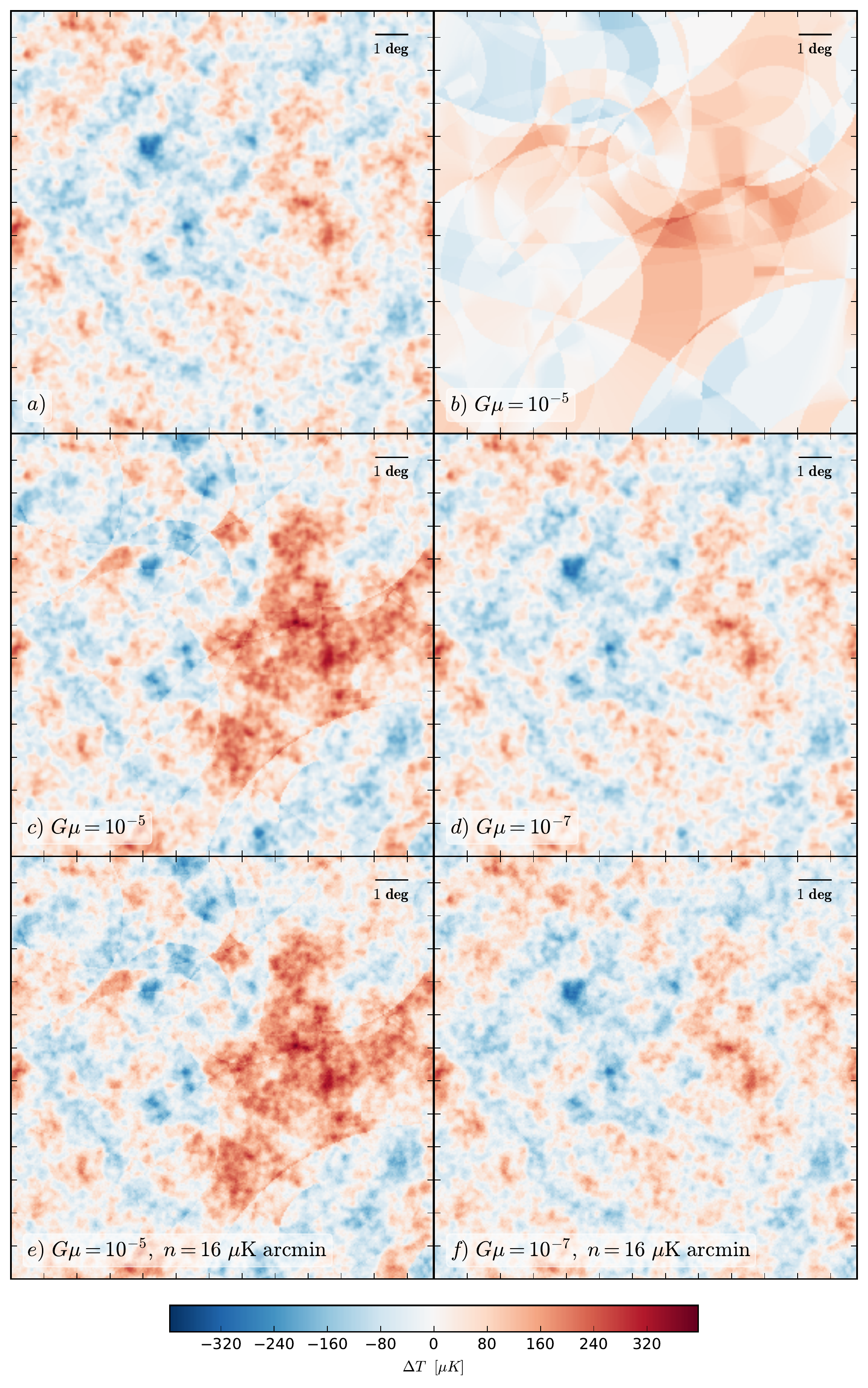}
\caption{Anisotropy maps of angular size $12.8^\circ \times 12.8^\circ$
and angular resolution of $1.5^\prime$. The top left panel comes from a 
particular realization of the $\Lambda$CDM
Gaussian CMB temperature map (no strings and no instrumental noise). 
The top right panel is a map from a simulation with only strings with a
tension given by $G \mu = 10^{-5}$. The two medium panels
are from simulations which include both Gaussian noise and
string signals, with the values of $G \mu$ indicated. With the
higher value of $G \mu$ the string signals are visible by eye,
but not with the lower value.
The bottom two panels show to which extent the string signals are
degraded when instrumental noise (at the level of the SPT-SZ
experiment) is included.The color indicates temperature.}
\label{maps}
\end{center}
\end{figure*}

The first map corresponds to what a vanilla $\Lambda$CDM cosmological
model would predict. The angular power spectrum $C_l$ was computed
using the CAMB code \cite{CAMB}. Like in \cite{Danos}, the Fourier space
temperature map was constructed from this according to
\be
\Delta T_G(k_x, k_y) \, = \, \sqrt\frac{C_l(k_x, k_y)}{2} \bigl(g_1(k_x, k_y) + i g_2(k_x, k_y) \bigr)
\, ,
\ee
where $g_1$ and $g_2$ are normally distributed random variables with mean zero
and variance one. The position space maps are then computed by taking the
inverse Fourier transform
\be
\Delta T_G(x, y) \, = \, iFFT[\Delta T_G(k_x, k_y)] \, ,
\ee
where $iFFT$ stands for the inverse Fourier transform. In order  to obtain
angular resolution of $1.5^\prime$ it is necessary to compute the angular
power spectrum up to beyond $l = 10'000$. The top left panel of Figure \ref{maps} shows a specific realization of the ``pure Gaussian'' map. The total angular size is $12.8^\circ \times 12.8^\circ$, and the angular resolution is $1.5^\prime$. 

Next we construct maps which correspond to
what a scaling distribution of long strings would produce. We adopt the
one-scale toy model of the distribution of long strings \cite{onescale}
described above. Furthermore, following what was done in \cite{Danos},
we divide the long strings into segments, and consider the segments
to persist for one Hubble time and to be uncorrelated on longer time
scales. In \cite{Danos}, the string segments at time $t$ were considered to
be straight lines with a fixed length $c_1 t$ with some constant $c_1$.
This procedure has two problems: firstly it does not take into account
the curvature of typical long strings (the mean curvature radius is $t$).
Secondly, the procedure introduces unphysical ends to the string
segments. Here we adopt a new procedure which overcomes these
two problems. We model the string segments as sections of circles,
and we connect the sections such as to avoid artificial ends.
Given the sky patch we simulate, the idea of the simulation is to
add up the temperature fluctuations by each string which contributes
to the temperature variations in this patch.

Specifically, we independently simulate a cosmic string map for time 
steps between the time of recombination until today
\be
t_{n + 1} \, = \, \alpha t_n
\ee
with $\alpha = 2$. For each time step we compute the corresponding
comoving Hubble distance corresponding to the curvature radius
of the strings:
\be
d_c(t_{n + 1}) \, = \, \alpha^{1/3} d_c(t_n) \, ,
\ee
where here $t_0 = t_{rec}$ is the time of recombination, the beginning of
the first simulation interval, and
\be
d_c(t_{rec}) \, \simeq \, 1.8^\circ \, .
\ee
At each time step we then randomly pick a first midpoint of a circular
string segment with a randomly picked angular size uniformly
distributed in the interval $[0, \pi]$. At the endpoints of
that string segment the next segments are placed, continued
until the string exits the map plus an additional Hubble distance. 
Two adjacent string segments are connected smoothly, i.e. such that there are no kinks. 
We then pick another midpoint and repeat the procedure until the total number of strings
per Hubble volume per time step is $N = 3$.

Given the string curvature, the strings will be moving towards
their center, thus creating a cooler patch inside the circle
segment and a hotter patch outside, according to the
equation
\be
\Delta T \, = \, 8 \pi G \mu \frac{{\tilde v} r}{2} T \, ,
\ee
where $T$ is the mean CMB background temperature,
${\tilde v} = 0.15$ (taken from the simulations of
\cite{Olum}) is the maximal string velocity factor
$v_s \gamma_s$, and $r$ is a random variable
uniformly distributed in the interval $[0 ,  1]$ to take into
account different velocities and projection effects.
The width of the temperature patterns are chosen to
be a fraction $f = 1/2$ of the respective Hubble radius.
Since the connecting point of the string segments cannot
move to the center of each circle at the same time,
it must have zero velocity. Hence, we linearly interpolate
${\tilde v}$ from one patch to the next over a fraction
of $20\%$ of the segment length.

The top right panel of Figure \ref{maps} 
shows the temperature map from one pure cosmic 
string simulation with a value of $G \mu = 10^{-5}$.
The angular extent and resolution are the same as
in the top left panel, as in all other panels of this
figure. The map represents a superposition
of the signals from strings in all Hubble time intervals.
The shorter edges are due to strings which influence
photons on our past light cone at early times, the longer
edges come from strings at later times. The color
indicates temperature, $G \mu$.

The combined maps of Gaussian and cosmic string
simulations are obtained via
\be
\Delta T_{total}(x, y) \, = \, \Delta T_G(x, y) + G \mu\ \Delta T_{CS}(x. y) \, ,
\ee
where the first term on the right hand side is what a
Planck-normalized $\Lambda$CDM model predicts,
and $\Delta T_{CS}$ is the contribution of cosmic strings
with $G \mu = 1$. Note that as $G \mu$ varies, the total
power in the anisotropy maps changes. But for the
values of $G \mu$ which we are interested in
($G \mu \lesssim 10^{-7}$) the difference in normalization 
is unimportant.

The middle left panel of Figure \ref{maps} 
shows the results of a simulation with $G \mu = 10^{-5}$.
In this map, the edges produced by the strings are still
visible by eye. But for the value $G \mu = 10^{-7}$, the
current upper bound on the string tension, the edges are
no longer visible, as shown in the middle right panel. 

Finally, we add to the joint cosmic string and Gaussian maps
a third map which corresponds to instrumental noise. We model this instrumental
noise as a pixel by pixel white noise contribution with fixed noise level
whose value is a free parameter and which can be chosen to match the
expected levels in various experiments.

The bottom two panels of Figure \ref{maps} show the resulting maps 
for the cosmic string parameters $G \mu = 10^{-5}$ (left panel)
and $G \mu = 10^{-7}$ (right panel) used before, with a noise level
corresponding to that of the SPT-SZ experiment.

In the following section we will show that by applying wavelet or curvelet transforms 
to these maps, the string signal can be retrieved and even made to stick out by eye.

\section{Wavelet and curvelet Analysis}

The wavelet transform (see e.g. \cite{wavelet} for an introduction) is a
method to analyze maps similar to Fourier analysis. In
Fourier analysis we expand the maps into basis functions which are
plane waves, and thus maximally delocalized in position space but
maximally localized in momentum space. Fourier analysis is very
useful in early universe cosmology to characterize fluctuations if
they are generated by a quantum vacuum process like in inflation.
In this case, the Fourier mode of each field satisfies a harmonic
oscillator equation and is quantized independently. The Fourier
spectrum hence contains the full information about the system
(modulo nonlinearities introduced through the dynamical evolution).
On the other hand, topological defects such as cosmic strings
produce maps which have special features in position space. These
features are washed out in a typical power spectrum based on
Fourier modes.

The wavelet analysis is based on decomposing the map into a set
of basis functions which are localized in both frequency and position space. Hence,
wavelets have the potential of yielding a better way to identify
cosmic string signals in cosmological maps.

 Like in the case of Fourier transformation in which the basis
 functions we expand in are obtained via the scaling of a basic
 plane wave, the wavelet transform uses a basis of functions
 obtained by scaling some ``mother'' function. In this work we
 used the Daubechies {\it db12} mother function \cite{Daubechies}. 
 We focus on the case of a two-dimensional map. Starting
 with the original image, we construct a first level {\it scaling image},
 which in the case of the simple Haar wavelet
 is a map with four times less pixels, the pixels
 representing the average of the values of the map of four
 neighboring pixels. In the case of the Haar wavelet,  
 the wavelet transform then takes the differences
 between these four pixels, yielding 
 three {\it details images} which represent the differences in
 the values of the map in horizontal, vertical or diagonal
 direction (the {\it horizontal details image}, {\it vertical details image}
 and {\it diagonal details image}, respectively). For
 the Daubechies {\it db12} wavelet which we use the
 geometrical interpretation of the four first level images
 is a bit more complicated. The full information
 of the original map is contained if we consider the scaling image
 plus the three details images. 
 The first level decomposition is also referred to as
 {\it finest scale}. The second level transform is
 then obtained by repeating the previous procedure and applying
 it to the scaling image.
 The computations were done using the Python package ``PyWavelets'' 
 (the module named {\it pywt}).
 
We next explored the potential of a curvelet analysis
to detect cosmic strings. Curvelets are defined through a partitioning of frequency space. One curvelet is obtained through rotation and translation of a mother curvelet that is the Fourier transform of a polar wedge in frequency space, defined by the support of a radial and an angular window with scale dependent window widths. Curvelets were introduced \cite{curvelet2000, Starck} to track discontinuities along edges and as such appear ideal to detect signatures of cosmic strings. Hence, curvelets should be even better suited to identify cosmic string signals than wavelets. In this work we have used Demanet's fast discrete curvelet
transform via edge wrapping for Matlab \cite{curvelet}. 

Both for the wavelet and the curvelet analysis, we will look at histograms of the pixel values in the finest scale details coefficients. From there we will perform kurtosis tests and Kolmogorov-Smirnov (KS) tests comparing the transforms of pure Gaussian maps to the combination of Gaussian and string maps.

%\begin{figure}[bt]
%\begin{center}
%\includegraphics[width=0.45\textwidth]{hergtl_fig2_cvlt_levels.pdf}
%\end{center}
%\caption{Comparison of the limits on the cosmic
%string tension which can be obtained using 
%curvelet analyses maps with SPT-3G specification,
%as a function of the level of the curvelet images used
%(all angles included). The horizontal
%axis is the string tension, the vertical axis gives the p-value
%from a KS test. The dotted horizontal line gives the $0.05$ p-value
%below which a detection of the string signal is considered significant. }
%\label{cvlt_levels}
%\end{figure}
%
%The resulting images are complex valued. For the purpose of 
%histograms the real and the imaginary images are concatenated. The curvelet images are
%labeled by level and by angle. The best results are obtained
%by making use of the two lowest level coefficients including all angles.
%Figure \ref{cvlt_levels} shows a comparison of how well a Kolmogorov-Smirnov (KS) test can
%extract a cosmic string signal for maps  with SPT-3G
%specification, as a function of the level of the curvelet images.
%The horizontal axis is the string tension, the vertical axis
%gives the p-value of the KS test. It is evident that Level 1
%gives stronger results than higher levels. By combining Levels 1 and 2,
% slightly tighter limits can be obtained than by using
%Level 1 maps alone.

\section{Results}
 
\begin{figure*}[bt]
\begin{center}
\includegraphics[width=0.6\textwidth]{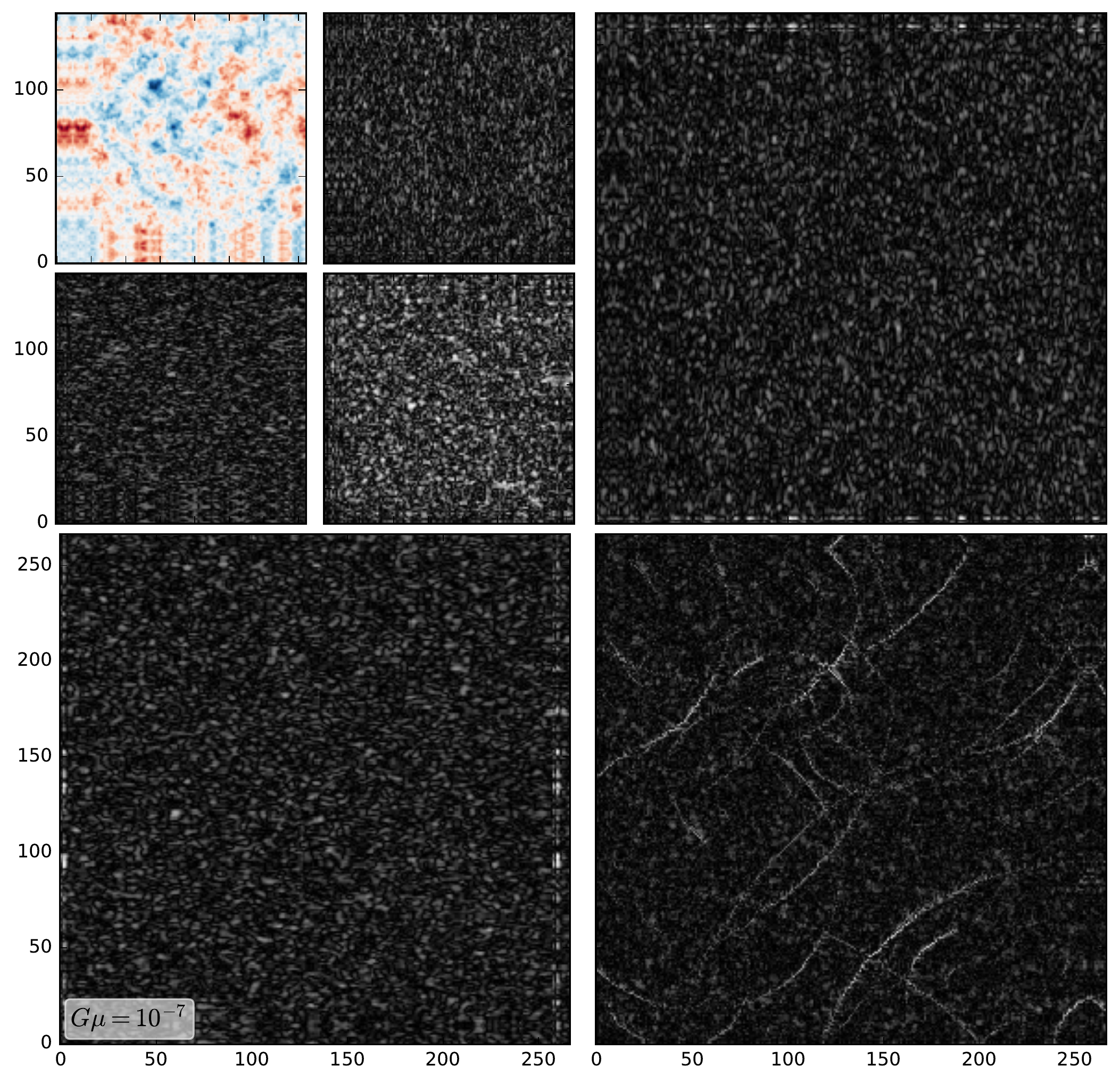}
\caption{Wavelet decomposition up to second level of the
combined CMB anisotropy map (without instrumental noise)
for a value of $G \mu = 10^{-7}$.
The bottom right panel gives the diagonal details image, the image
which is best for finding string signals. The other panels are
described in the main text. The string signal is clearly visible
by eye in the diagonal details image even though it is not visible
by eye in the original map.}
\label{wvlt_details}
\end{center}
\end{figure*}

 We first discuss the results of our wavelet analysis.
 Figure \ref{wvlt_details} shows the wavelet decomposition of our maps (without
 instrumental noise) up
 to the second level in the case of combined Gaussian / string
 fluctuations with a value of $G \mu = 10^{-7}$. The top left 
 quadrant shows the second level scaling and details images, 
 the top right quadrant shows the first level horizontal details image, 
 the bottom left the first level vertical details image, the bottom 
 right the diagonal details image. 
 For this value of $G \mu$, the string signals were not visible
 by eye in the original map, but they are clearly visible in the diagonal
 details image at first level. This demonstrates the promise that the wavelet method has at identifying string signals. 

\begin{figure*}[bt]
\begin{center}
\includegraphics[width=\textwidth]{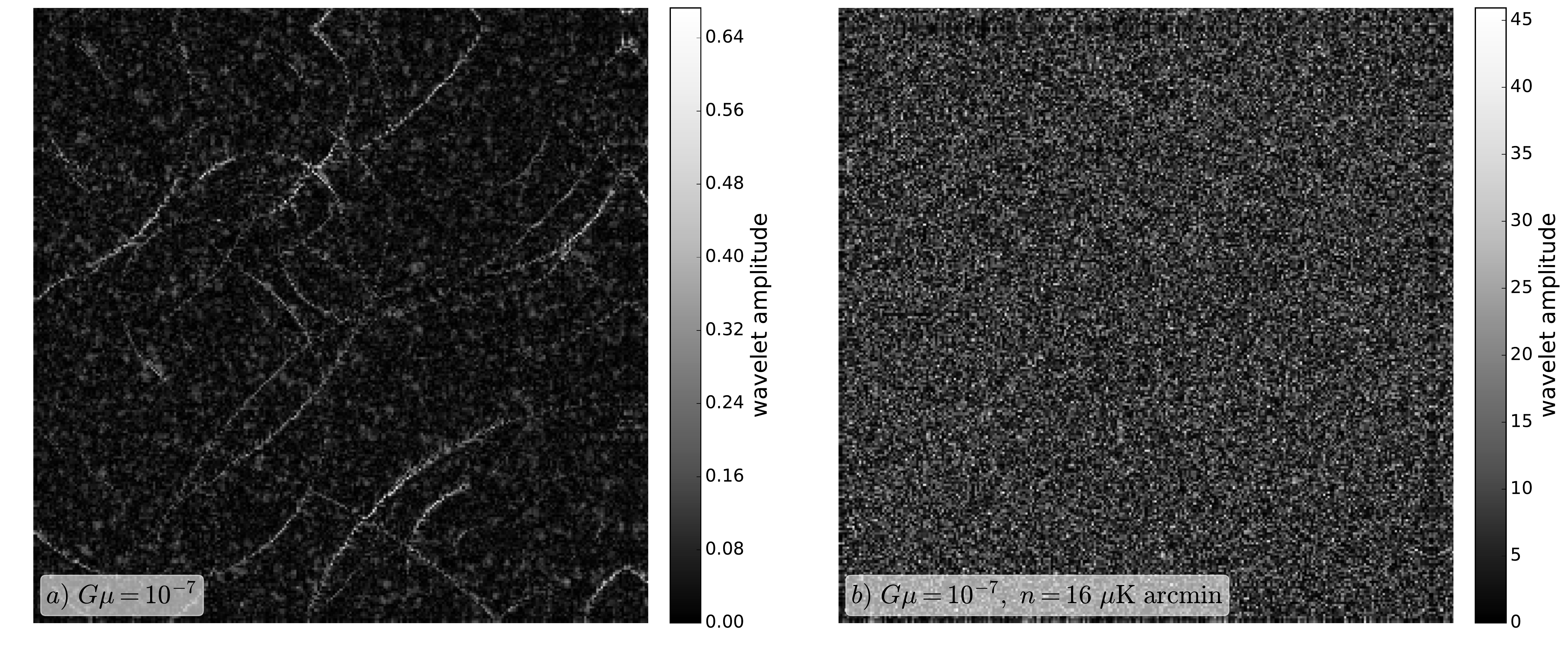}
\caption{Diagonal details image at first level for a CMB anisotropy map
with and without instrumental noise and for the string tension
given by $G \mu = 10^{-7}$. The left panel is without instrumental
noise, and the string signals are visible by eye. In the right
panel (which includes instrumental noise at the level of the
SPT-SZ experiment) the string signal is no longer visible by eye.}
\label{wvlt_diag}
\end{center}
\end{figure*}

The panels of Figure \ref{wvlt_diag} show the first level diagonal details images for the
value $G \mu = 10^{-7}$ without (left panel) and with (right panel)
instrumental noise. Without noise, the string signals are  visible by eye,
but with noise (noise levels of the SPT-SZ experiment) they are no
longer visible.

\begin{table}[b]
\centering
\def\arraystretch{1.3}
\setlength{\tabcolsep}{6pt}
\begin{tabular}{ r c r r }
\toprule
& \multicolumn{1}{c}{$G\mu$}    & \multicolumn{1}{c}{$k$} & \multicolumn{1}{c}{$p_k$} \\
\colrule
Gaussian map    &                       & $-0.011$ & $0.561$   \\
CS map          & (all)                 & $35.702$ & $ < 0.001$ \\
\\
combined maps   & $1\times10^{-7}$      & $3.217$  & $< 0.001$ \\
                & $5\times10^{-8}$      & $0.327$  & $< 0.001$ \\
                & $4\times10^{-8}$      & $0.137$  & $< 0.001$ \\
                & $3\times10^{-8}$      & $0.037$  & $0.048$ \\
                & $2\times10^{-8}$      & $-0.003$ & $0.883$ \\
                & $1\times10^{-8}$      & $-0.012$ & $0.537$ \\
\botrule
\end{tabular}
\caption[Kurtosis from wavelet decomposition]{Excess kurtosis $k$ from the diagonal 
        details image of the 1st level wavelet decomposition  of one map and p-value $p_k$ from the 
        kurtosis test for different values of the string tension $G\mu$. (without white noise)}
\label{tab:wvlt}
\end{table}

\begin{figure*}[bt]
\begin{center}
\includegraphics[width=0.75\textwidth]{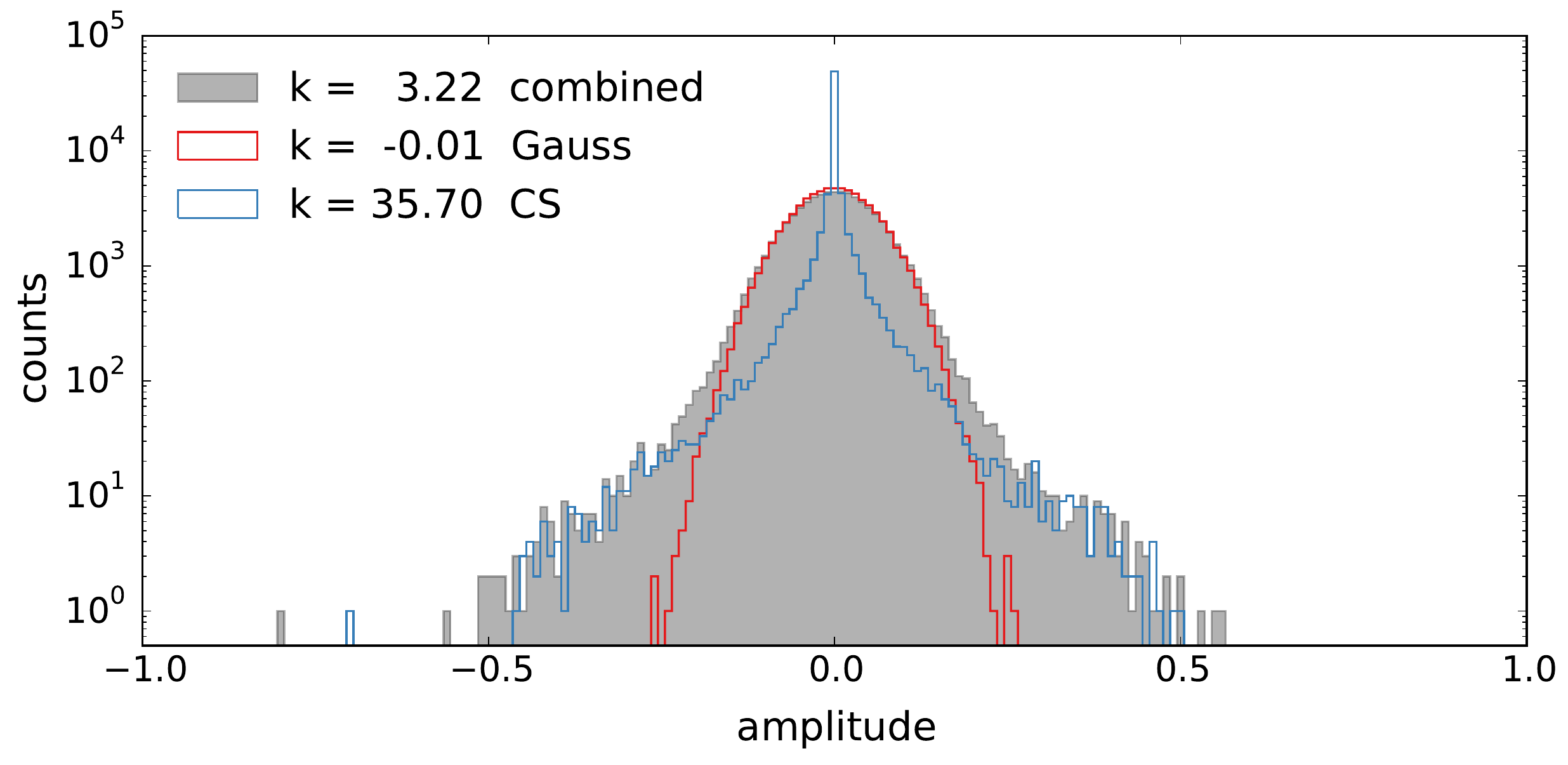}
\caption{Histogram of the values in the diagonal details image of the wavelet decomposition of the map with Gaussian fluctuations plus strings with $G \mu = 10^{-7}$ but without instrumental noise (Figure \ref{wvlt_diag}a). The kurtosis due to the large map values along the strings is visible by eye.}
\label{wvlt_hist}
\end{center}
\end{figure*}

Let us, for the moment, turn off instrumental noise. Then, 
if we further reduce the value of $G \mu$ from that used in the left
panel of Figure \ref{wvlt_diag}, then the string signals
eventually are no longer visible by eye in the first level diagonal
details image. However, with statistical analyses of this details
map we can probe deeper. We have explored a very simple
statistic, namely the kurtosis of the histogram of the values of the map.
The histogram of a Gaussian map is Gaussian and hence its excess
kurtosis is zero. The diagonal details map of a pure string map is,
on the other hand, characterized by a few points with a very large value
while most points have very small amplitude (the large amplitude
points are along the strings). Hence, the map has a large kurtosis.
The combined map will have a non-vanishing kurtosis coming from
the contributions of the strings. Figure \ref{wvlt_hist} shows the histogram of
the diagonal details map for the  case $G \mu = 10^{-7}$
 (once again without instrumental noise).
The non-Gaussian contribution due to strings is visible by eye for
large absolute values. In Table \ref{tab:wvlt} we give the value of the kurtosis $k$
and the related p-value $p_k$ under the hypothesis
that the kurtosis is consistent with a
Gaussian distribution for various values of $G \mu$ in the range
between $10^{-8}$ and $10^{-7}$. The bottom line is that,  in the absence
of instrumental noise, with a
false probability of less than $5 \%$, strings could be identified
down to a value of $G \mu = 3 \times 10^{-8}$, a factor of $3$
below the current limits.

However, adding pixel by pixel white noise to mimic instrumental
noise degrades the results substantially. The reason is that the
white noise has most power on the small scales where in position
space the cosmic string signal is concentrated. The right
panel of Figure \ref{wvlt_diag} shows the wavelet diagonal
details image for the same parameters as used in the left panel, 
but including white noise with the level of the SPT-SZ
experiment. As is evident, the string signal is
now no longer visible.

\begin{figure*}[bt]
\begin{center}
\includegraphics[width=\textwidth]{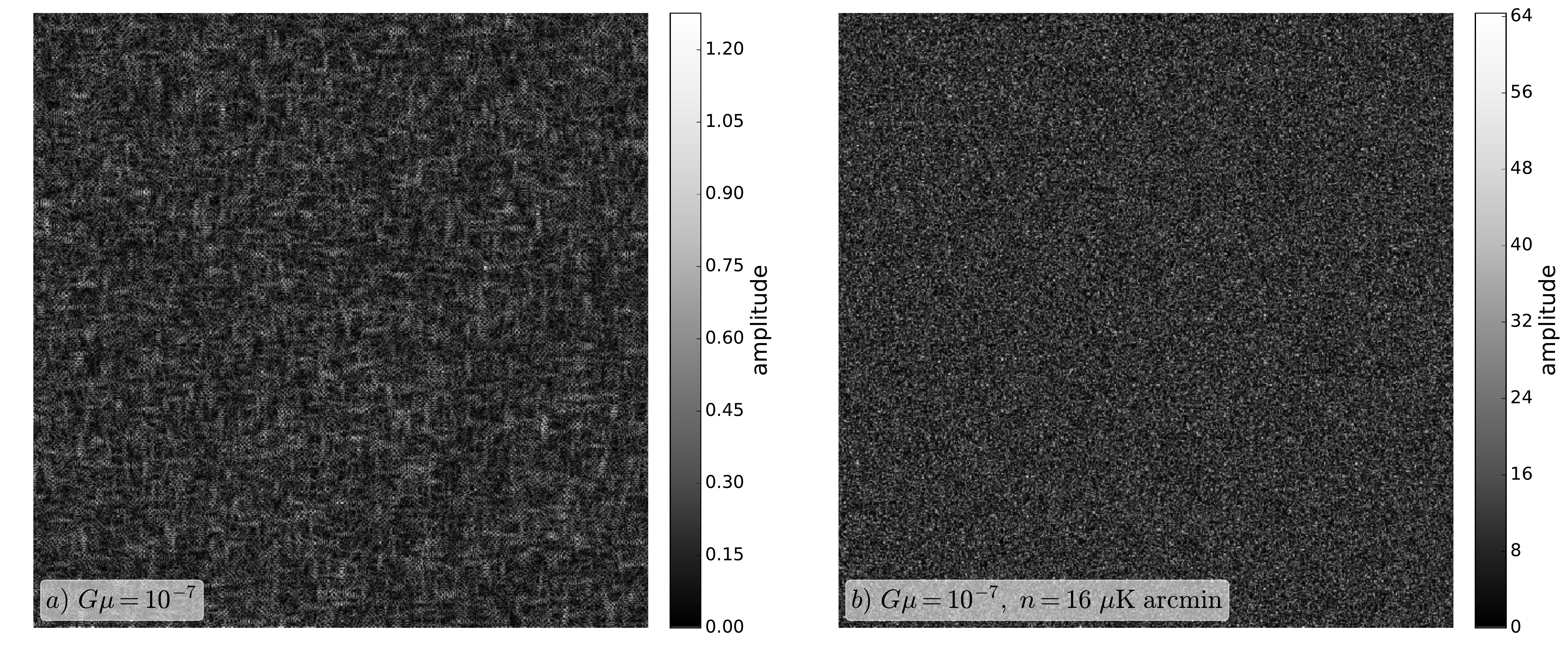}
\caption{Reconstruction of the maps from finest scale curvelet images.
The left panel corresponds to a CMB anisotropy map with both
Gaussian fluctuations and cosmic strings with $G \mu = 10^{-7}$, but
without instrumental noise. Here the string signal still are visible by eye. On the right panel, instrumental noise
at the level of the SPT-SZ experiment is included and the string signal is no longer visible by eye.}
\label{cvlt_rec}
\end{center}
\end{figure*}

Next we turn to the results of the curvelet analysis.
The panels in Figure \ref{cvlt_rec} show the curvelet reconstruction of the map from finest
scale coefficients. The left panel corresponds to a simulation with both
Gaussian noise and cosmic strings with a tension of  $G \mu = 10^{-7}$ but
without instrumental white noise.  
The strings are visible by eye. 
As in the case of wavelets, we look at histograms of the pixel amplitude of the coefficient images, where the stringy features lead to a large kurtosis. If the
string tension is lowered further, the string signals are no longer
visible by eye, but they remain identifiable via the kurtosis of the histogram,
namely an excess in the number of pixels with large absolute values which
correspond to points lying along the strings.

In Table \ref{tab:cvlt} we give analogously to Table \ref{tab:wvlt} the value of the kurtosis $k$ and the associated
probability $p_k$ that the result could be from a Gaussian distribution,
for values of $G \mu$ in the interval $[10^{-8}, 10^{-7}]$ (in the
absence of white noise). We see
that the probability for deviations from a Gaussian distribution is
significant down to a value of $G \mu = 3 \times 10^{-8}$.

\begin{table}[b]
\def\arraystretch{1.3}
\setlength{\tabcolsep}{6pt}
\centering
\begin{tabular}{ r c r r }
\toprule
& \multicolumn{1}{c}{$G\mu$}    & \multicolumn{1}{c}{$k$} & \multicolumn{1}{c}{$p_k$} \\
\colrule
Gaussian map    &                       & $0.055$   & $0.068$   \\
CS map                  & (all)         & $107.267$ & $< 0.001$ \\
\\
combined maps   & $1\times10^{-7}$      & $2.199$   & $< 0.001$ \\
                & $5\times10^{-8}$      & $0.204$   & $< 0.001$ \\
                & $4\times10^{-8}$      & $0.110$   & $0.001$ \\
                & $3\times10^{-8}$      & $0.066$   & $0.036$ \\
                & $2\times10^{-8}$      & $0.051$   & $0.105$ \\
                & $1\times10^{-8}$      & $0.051$   & $0.101$ \\
\botrule
\end{tabular}
\caption[Kurtosis from curvelet decomposition]{Excess kurtosis $k$ from the 
        first angle curvelet decomposition and p-value $p_k$ from the kurtosis test for 
        different values of the string tension $G\mu$. (without white noise)}
\label{tab:cvlt}
\end{table}

Once again, adding instrumental noise degrades the
power of the curvelet method. In the right panel of Figure \ref{cvlt_rec}
we show the CMB anisotropy map reconstructed from finest scale 
curvelets for the same cosmic string parameter as in the left panel,
but this time including instrumental white noise with the
amplitude of the SPT-SZ experiment. The string signal has been washed out.

\begin{figure}[b]
\begin{center}
\includegraphics[width=0.45\textwidth]{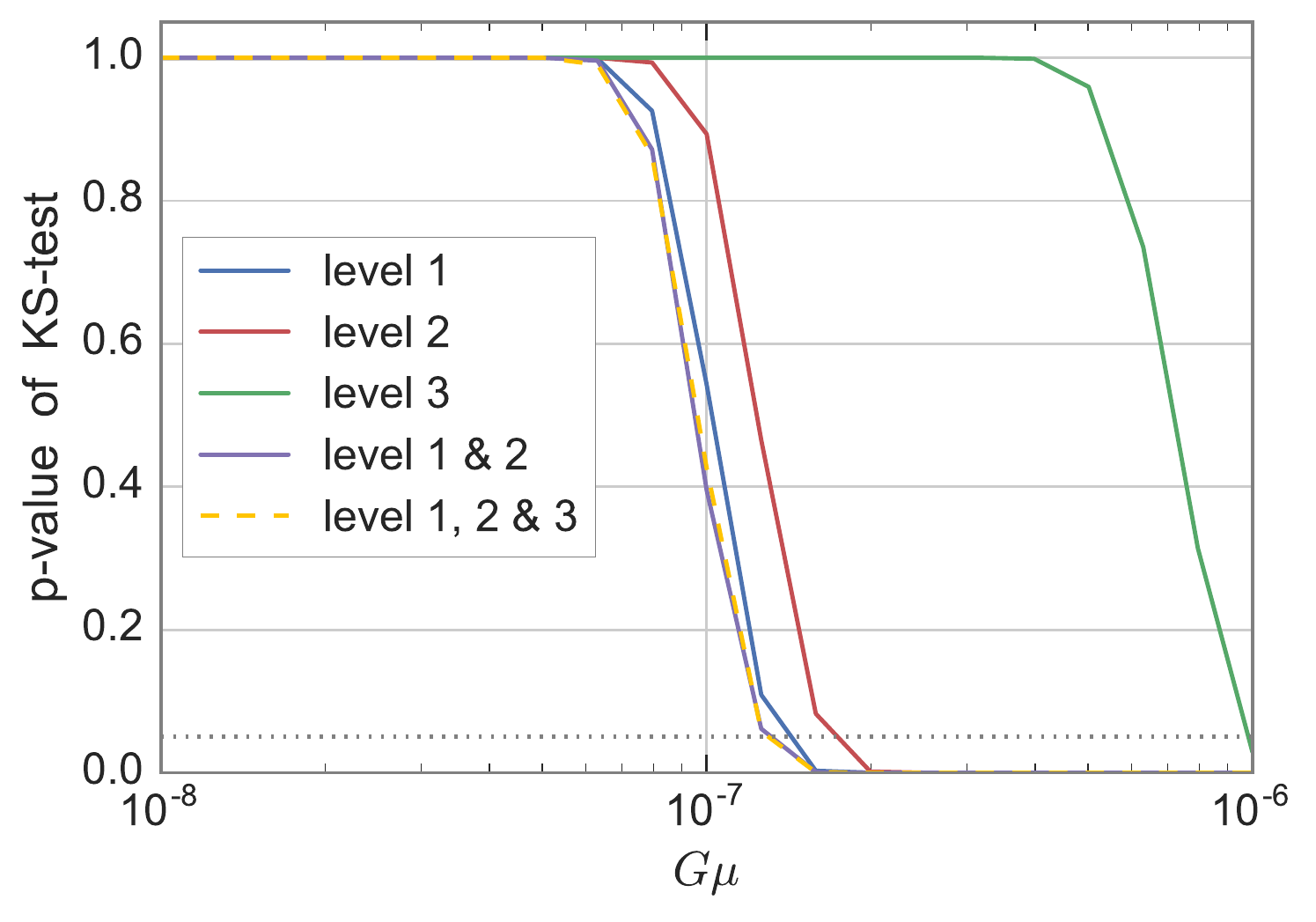}
\end{center}
\caption{Comparison of the limits on the cosmic
string tension which can be obtained using 
curvelet analyses maps with SPT-3G specification,
as a function of the level of the curvelet images used
(all angles included). The horizontal
axis is the string tension, the vertical axis gives the p-value
from a KS test. The dotted horizontal line gives the $0.05$ p-value
below which a detection of the string signal is considered significant. }
\label{cvlt_levels}
\end{figure}

We have varied the number of curvelet coefficient angels and scales taken into account when constructing the histograms. Since the resulting images are complex valued, we concatenated the real and imaginary images. 
The best results are obtained by making use of the first two level coefficients including all angles.
Figure \ref{cvlt_levels} shows a comparison of how well a Kolmogorov-Smirnov (KS) test can
extract a cosmic string signal for maps  with SPT-3G
specification, as a function of the level of the curvelet images.
The horizontal axis is the string tension, the vertical axis
gives the p-value of the KS test. It is evident that Level 1
gives stronger results than higher levels. By combining Levels 1 and 2,
 slightly tighter limits can be obtained than by using
Level 1 maps alone.

\begin{figure*}[t]
\begin{center}
\includegraphics[width=0.75\textwidth]{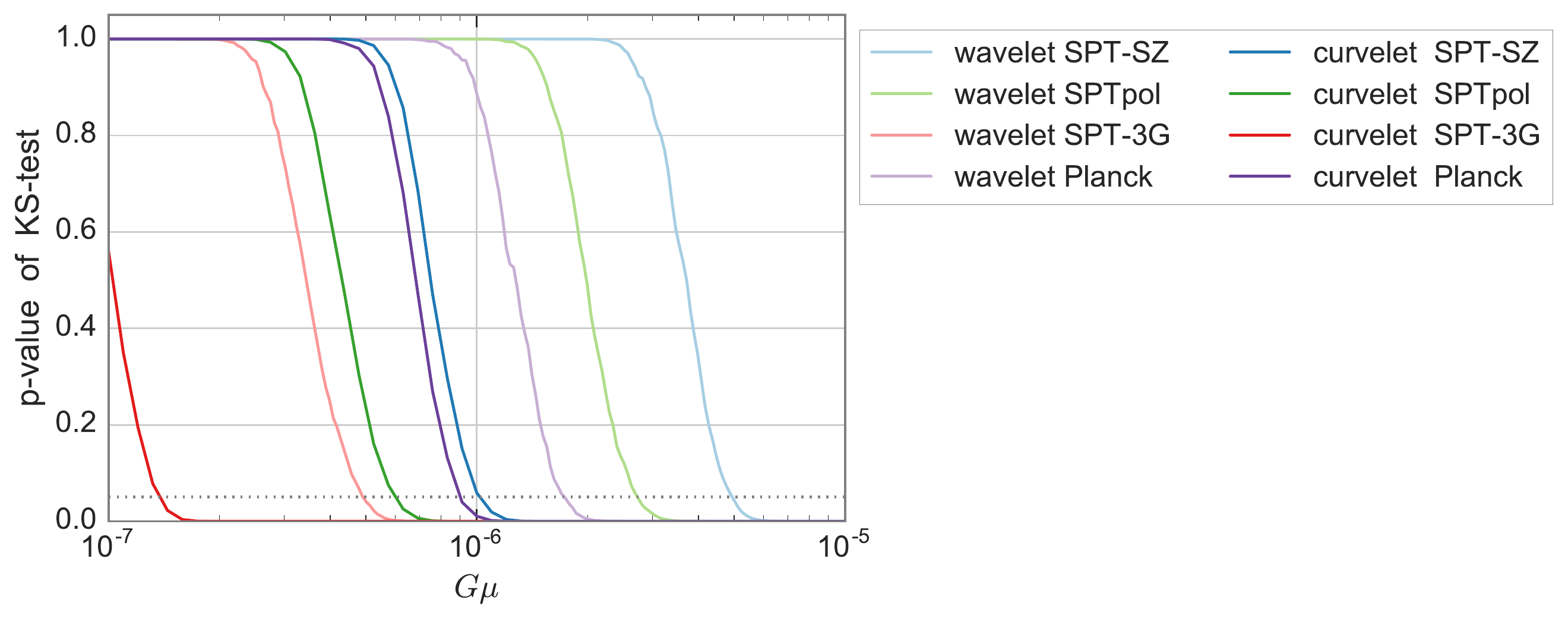}
\caption{ Comparison of the limits on the cosmic
string tension which can be obtained using wavelet and
curvelet analyses of SPT and Planck maps. The horizontal
axis is the string tension, the vertical axis gives the p-value
from a KS test. The dashed horizontal line gives the p-value
below which a detection of the string signal is significant. The
wavelet analysis made use of the finest scale diagonal details image,
the curvelet analysis made use of the first two levels and all angles.}
\label{ks_exp}
\end{center}
\end{figure*}

Finally, we compared the limits on the cosmic string tension
which can be obtained from various ongoing experiments, specifically
three stages of the South Pole Telescope (SPT) experiment - 
SPT-SZ \cite{SPT-SZ}, SPTpol \cite{SPTpol} and SPT-3G \cite{SPT3G} -
and the Planck survey \cite{Planck} \footnote{For Planck we have
used the properties of the $143~\text{GHz}$, for SPT the $150~\text{GHz}$ channel.}. We constructed maps with
size, angular resolution and instrumental noise appropriate to
these surveys (the parameters used are given in Table \ref{tab:exp}). We
computed the wavelet and curvelet transforms of maps with and
without cosmic strings (in addition to Gaussian primordial fluctuations and white noise), and performed a KS analysis for the p-value for which the string signal is visible as
a function of the cosmic string tension $G \mu$. In the case
of the wavelet analysis, we used  the finest scale
diagonal details images, for the curvelet analysis
we used the first two levels and all angles. The results are given in Figure \ref{ks_exp}
and Table \ref{tab:exp}.

We see that the curvelet analysis provides stronger limits than
the wavelet analysis. This result is easy to understand since
curvelets are better adapted to linear signals than wavelets, and thus more robust to the added white noise. We
find that, at the current noise levels, the Planck maps yield
tighter constraints than the SPT-SZ maps, comparable
constraints as the SPTpol maps, but that
the third generation SPT experiment will be able to provide
more stringent constraints than Planck because of the
better angular resolution and the reduced noise levels.

\begin{table*}[bt]
\centering
\def\arraystretch{1.4}
\setlength{\tabcolsep}{10pt}
\begin{tabular}{ l D{.}{.}{3.0} D{.}{.}{5.0} D{.}{.}{1.1} D{.}{.}{2.1} S[table-format = 1.1e3] S[table-format = 1.1e3] }
\toprule
\multicolumn{1}{l}{experiment} & \multicolumn{1}{c}{frequency}	& \multicolumn{1}{c}{survey area} & \multicolumn{1}{c}{resolution} & \multicolumn{1}{c}{sensitivity} & \multicolumn{2}{c}{$G\mu$ at 95\%}  \\
&\multicolumn{1}{c}{$\mathrm{[GHz]}$}&\multicolumn{1}{c}{$\mathrm{[deg^2]}$}&\multicolumn{1}{c}{$\mathrm{[arcmin]}$}&\multicolumn{1}{c}{$\mathrm{[\mu K\ arcmin]}$}& \multicolumn{1}{c}{wavelets} & \multicolumn{1}{c}{curvelets}  \\
\colrule
SPT-SZ & 150 & 2500   & 1  & 16   & 5e-6   & 1e-6 \\
SPTpol & 150 & 500    & 1  & 6    & 3e-6   & 6e-7 \\
SPT-3G & 150 & 2500   & 1  & 1.6  & 5e-7   & 1.4e-7 \\
Planck & 143 & 32029  & 7.3& 33   & 1.8e-6 & 9e-7 \\
\botrule
\end{tabular}
\caption{Parameters of various experiments from \cite{PlanckHFI, SPT3G} used for map generation, and the corresponding $G\mu$ limits (at 95\% confidence).}
\label{tab:exp}
\end{table*}

\section{Conclusions and Discussion}

In this paper we have explored the potential of extracting signals
of cosmic strings from CMB temperature maps using wavelets
and curvelets. We have constructed temperature maps of
the size and angular resolution corresponding to various
ongoing CMB experiments containing string signals,
Gaussian primordial perturbations, and instrumental noise
modelled as white noise, and we have used a KS test
on maps with and without string signals to find the level
of the string tension for which the string signals can
be extracted in a statistically significant way.
 
We find that the curvelet analysis 
yields stronger constraints than an analysis based on
wavelets, especially in the presence of white noise. Wavelets, in turn, yield stronger constraints
than what can be obtained using the Canny edge detection algorithm. 
We find that string signals can be seen down to a value 
$G \mu = 1.4\times 10^{-7}$ in maps with the specification of the
upcoming SPT-3G experiment. This is significantly stronger
than corresponding bounds from Planck maps. The
improvement is due to the better angular resolution.

Bounds similar to the one we have obtained also result from
CMB angular power spectrum analyses \cite{Dvorkin, PlanckCS}.
In these analyses, the effect of string-induced fluctuations
on the pattern of acoustic oscillations in the CMB is studied.
There are other effects which can mimic the effect of
cosmic strings on the angular power spectrum (see e.g.
\cite{Turok}). An advantage of our method is that we are
looking for signals distinctive to strings. In fact, we have
seen that without instrumental noise the strings would
be directly visible by eye in wavelet and curvelet images down
to a value of $G \mu = 5\times 10^{-8}$.

Applied to maps with Planck specification, our limits are
comparable to what is obtained in \cite{PlanckCS}. The
analyses, however, have complementary strengths and
weaknesses. Our analysis focuses on idealized string
signals, and we are able to freely vary parameters such
as the number $N$ of strings per Hubble volume, and
the mean string velocity, parameters which depend on the
type of cosmic string evolution simulation being used and
on which there is no agreement, while the analysis of
\cite{PlanckCS} uses maps which result from a specific
cosmic string simulation. On the other hand, the analysis
of \cite{PlanckCS} is based on less idealized maps which
contain the effects not only of the long string segments, but
also of string loops and of small-scale structure on the
long strings. An obvious advantage of our method is that
it is more easily applicable to a range of different experimental
designs.

We have only explored one type of mother function for
wavelets. There is a lot of room for exploring
possible improvement of the detection algorithm by
optimizing the mother functions chosen.

We have neglected the small-scale structure (see e.g. \cite{small} for some recent
studies of small-scale structure on strings) which is expected to build up
on the sides of long strings. We have modeled the effects of the
small scale structure by taking a narrow band around the
temperature discontinuity across the string and replacing
$\delta T$ by $0$ in this band. In that case, we find that the limit on $G \mu$ to which strings can
be detected deteriorates by a factor of $2$.

A potential followup project would be to apply our
wavelet and curvelet algorithms to CMB temperature
anisotropy maps constructed from numerical cosmic
string evolution simulations (see \cite{Ringeval, CSsimuls} for a partial
list of papers discussing such simulations).

The results of this work encourage us to consider application
of wavelet and curvelet statistics to other observational windows,
e.g. CMB polarization maps, or three-dimensional 21cm redshift
surveys.

\section*{Acknowledgement}
\noindent

One of us (RB) wishes to thank the Institute for Theoretical Studies of the ETH
Z\"urich for kind hospitality. RB acknowledges financial support from Dr. Max
R\"ossler, the ``Walter Haefner Foundation'' and the ETH Zurich Foundation, and
from a Simons Foundation fellowship. The research is also supported in
part by funds from NSERC and the Canada Research Chair program.

\end{document}